\documentclass[12pt]{iopart}
\usepackage[dvips]{graphicx}
\usepackage[dvips]{color}
\usepackage{iopams}

\definecolor{MyColor}{rgb}{0.0,0.4,1} 
\def\m#1{\mathbf{#1}}
\def\an{\alpha_n}

\def\bn{\beta_n}

\def\Jd{J^{(d)}}
\def\Js{J^{(s)}}

\def\e{\varepsilon}
\def\wrt{{\em{w.r.t.}} }
\def\hot{{\rm{h.o.t.}} }
\def\kB{k_\mathrm{B}}
\def\mR{\mathbf{R}}
\def\mS{\mathbf{S}}
\def\mV{\mathbf{V}}
\def\mU{\mathbf{U}}
\def\mY{\mathbf{Y}}
\def\mZ{\mathbf{Z}}
\def\mC{\mathbf{C}}
\def\mG{\mathbf{G}}
\def\mW{\mathbf{W}}

\def\Lop{\mathcal{L}}

\let\rho=\varrho

\newcommand{\pderiv}[1]{\frac{\partial #1}{\partial t}}
\newcommand{\cor}[1]{\langle #1\rangle}

\begin{document}

\title[Stochastic model of anomalous heat
  transport]{A stochastic model of anomalous heat
  transport: analytical solution of the steady state}
\author{S Lepri, C Mej{\'{\i}}a-Monasterio, and A Politi}
\address{Istituto dei Sistemi Complessi, Consiglio Nazionale
delle Ricerche, via Madonna del Piano 10, I-50019 Sesto Fiorentino, Italy}
\eads{\mailto{carlos.mejia@fi.isc.cnr.it}, \mailto{stefano.lepri@isc.cnr.it}, \mailto{antonio.politi@isc.cnr.it}}
\begin{abstract}
We consider a one-dimensional  harmonic crystal  with conservative  noise, in
contact with two stochastic Langevin heat  baths at different temperatures.
The  noise term  consists of collisions  between  neighbouring oscillators that
exchange their momenta, with a rate $\gamma$. The stationary equations for the
covariance matrix are exactly solved in the thermodynamic limit ($N\to\infty$).
In particular, we derive an analytical expression for the temperature
profile, which turns out to be independent of $\gamma$. Moreover, we obtain
an exact expression for the leading term of the energy current, which scales
as $1/\sqrt{\gamma N}$. Our theoretical results are finally found to be
consistent with the numerical solutions of the covariance matrix for finite $N$.
\end{abstract}
\submitto{\JPA}

\section{Introduction}
\label{sec:intro}

Understanding the statistical properties of open, many-particles system is one
of the challenges of nonequilibrium statistical mechanics.  From a fundamental
point  of view,  a successful  approach would  require to  find,  and possibly
compute  explicitely, a statistical  measure for  (at least)  systems steadily
kept out of  equilibrium.  Some insight has been gained  over the years mostly
thanks  to  the  analysis  of  specific  models  (for  a  recent  account  see
e.g. \cite{Bertini07} and references therein).   A related open problem is the
derivation of  phenomenological transport laws from  the microscopic dynamics,
without any \textit{ad hoc} statistical  assumption.  An example is the famous
law, postulated by  Joseph Fourier almost two hundred  years ago, relating the
heat  flux  $J$ flowing  within  a solid  material  to  the local  temperature
gradient,
\begin{equation} \label{eq:fourier}
J = -\kappa \nabla T \ ,
\end{equation}
where the  constant of proportionality  $\kappa$, is the  thermal conductivity.

In the lack of a general  framework, simple models are precious to attack such
difficult problems \cite{LLP03,BLRB00}. An  instance, dating back to 1967, was
provided  by Rieder, Lebowitz  and Lieb  who considered  heat conduction  in a
chain of  harmonic oscillators connected  at its boundaries to  two stochastic
heat  baths   \cite{RLL67}.  They  showed   that  the  invariant   measure  in
phase--space  (i.e. the  stationary solution  of the  associated Fokker-Planck
equation) is  a multivariate Gaussian.  Furthermore, they proved that,  due to
the integrability  of the  underlying dynamics,  such a model  is not  able to
support a temperature gradient.  However, this  is one of the very few systems
that  have   been  rigorously  solved.    Extensions  of  this   model,  where
anharmonicities are introduced by  means of self-consistent local thermostats,
were  extensively studied \cite{BRV70,BLL04,DR06}.   In recent  years, further
attempts to derive  Fourier's law in deterministic systems  have been reported
\cite{LS06,BK07,EY06,GG08}.

As a complementary approach, stochastic models have played an important r\^ole
in understanding how energy is microscopically transported. This is mainly due
to the  fact that the stochastic  approach seems to easily  yield results that
would require much more efforts  by adopting the dynamical approach.  In fact,
while   stochastic  models   are  assumed   to  be   a   reduced  (mesoscopic)
representation of the ``chaotic'' microscopic dynamics, they are free from the
intricacies  of  the fractal  structures  arising  in deterministic  dynamics.
Actually,  the  leap   from  such  class  of  models   to  even  the  simplest
deterministic,   nonlinear  ones  is   still  a   challenge  for   the  theory
\cite{BLRB00}.  At  the simplest  level of modeling,  energy is assumed  to be
randomly    exchanged    between    neighbouring    sites   of    a    lattice
\cite{Davies78,KMP82,EPRB99,GKR07}. This  class of systems  has the invaluable
advantage  of  allowing for  a  mathematically  rigorous  treatment, which  is
usually unfeasible  in the deterministic case.  Recently,  systems of harmonic
oscillators exchanging  energy with ``conservative" noise have  been proven to
admit a unique stationary state consistent with \eref{eq:fourier} \cite{BO05}.
However, if the  additional constraint that the random  process conserves also
linear  momentum  is   imposed,  the  equilibrium  energy-current  correlation
function decays as $t^{-d/2}$ ($d$  being the lattice dimension) and transport
becomes anomalous in $d\le 2$ \cite{BBO06}.  This means that \eref{eq:fourier}
breaks  down  as  $\kappa$ diverges  with  the  system  size. The  results  of
\cite{BBO06}  thus provide  a  rigorous  basis to  the  numerical evidence  of
anomalous  transport  and diffusion  in  deterministic  nonlinear models  with
momentum  conservation \cite{LLP03}, with  the only  exception of  the coupled
rotor chain \cite{giardina00,gendelman00}.

In  this paper  we  consider the  problem of  heat  conduction in  a chain  of
harmonic  oscillators, coupled  at  its boundaries  with  two stochastic  heat
baths.  In  addition  to  the  deterministic  bulk  dynamics,  we  consider  a
stochastic ``noise''  dynamical term, consisting of collisions  occurring at a
given rate $\gamma$, that exchange the momenta of a stochastically chosen pair
of  neighbour  oscillators.  The   stochastic  contribution  to  the  dynamics
maintains the linearity of the associated Fokker-Planck equation.

Recently, following a principal component analysis, we  have numerically found
that, in the basis identified by the eigenvectors of the covariance matrix, the
nonequilibrium invariant measure of this model can be effectively expressed as
the product of independent distributions aligned along collective modes that
are spatially localized with power-law  tails \cite{DLLP08}. Moreover, several 
variables, such as the amplitudes of these modes, turn out to be Gaussian 
distributed. Accordingly, it appears that the unavoidable deviations from a
Gaussian behaviour are confined to not-so-relevant observables, so that the 
nonequilibrium invariant measure can be effectively considered to be a 
multivariate Gaussian. Within this aproximation, the covariance matrix provides
a complete description of the corresponding invariant measure. Here,
with  the  help  of  a  suitable  continuum limit,  we derive leading  order
expressions for the covariance matrix in the steady nonequilibrium state, from
which explicit formulae for the temperature profile and the  energy current,
are  obtained. It should  be noticed  that this  is the  first example  of an
analytic expression for  the temperature profile in a  system characterized by
anomalous heat transport (i.e.  diverging conductivity).

The paper  is organized as  follows. In Section \ref{sec:model},  we introduce
the  stochastic model.  In  Section \ref{sec:cov},  we  define the  covariance
matrix $\mC$ and  write down the coupled equations  governing the evolution of
$\mC$ towards  its stationary  value. The  key results of  the paper  are also
summarized  there.  Section~\ref{sec:perturb}  contains  the  details  of  the
analytical   calculation  of   the   stationary  covariance   matrix  in   the
thermodynamic  limit  $N\to \infty$.  Our  approach  is  based on  a  suitable
continuum  approximation, in  which  the finite-difference  equations for  the
entries  of $\mC$  are  replaced  by partial  differential  equations for  the
corresponding field-like variables. We obtain the covariance matrix to leading
order    in    the    smallness   parameter    $\varepsilon=1/\sqrt{N}$.    In
Section~\ref{sec:concl}  we  discuss  the  physical meaning  of  the  analytic
expressions, compare them  with the numerical solution for  finite size chains
and briefly comment on the open problems.

\section{Stochastic Model}
\label{sec:model}

We consider a  homogeneous harmonic chain of $N$ oscillators  of unit mass and
frequency  $\omega$, in contact  with two  different stochastic  Langevin heat
baths at its  extrema and fixed boundary conditions. The  dynamics in the bulk
of the chain is governed by the Hamiltonian
\begin{equation} \label{eq:H}
H(\vec{q},\vec{p},t) \ = \ \sum_{i=1}^{N} \left[ \frac{p_i^2}{2} +
\frac{\omega^2}{2}\left(q_{i+1} - q_{i}\right)^2 \right] \,
\end{equation}
Furthermore, the  $1$-st and $N$-th  oscillators are coupled to  Langevin heat
baths at temperatures $T_\pm=T\pm\Delta  T/2$ respectively ($T$ is the average
temperature $(T_++T_-)/2$).  Then the equations of motion become
\begin{equation} \label{eq:eqs-motion}
\begin{array}{rcl}
\dot q_n & \ = \ & p_n \\
\dot p_n & \ = \ & \omega^2 (q_{n+1} - 2q_n + q_{n-1}) + 
 \delta_{n,1}(\xi_+ - \lambda \dot q_1) + \delta_{n,N}(\xi_- -\lambda \dot
 q_N)  \ ,
\end{array} 
\end{equation}
where $\xi_-$ and $\xi_+$ are  independent Wiener processes with zero mean and
variance  $2\lambda  k_BT_-$ and  $2\lambda  k_BT_+$  respectively. The  fixed
boundary conditions are enforced by  setting $q_0 = q_{N+1}= 0$.  In addition,
the chain undergoes random binary collisions, at a rate $\gamma$, in which the
momenta  of a  couple of  neighbouring oscillators  are exchanged.   Thus, the
resulting dynamics conserves both total momentum and energy.

The phase space probability density $P(\vec{q},\vec{p},t)$ of this model, is a
solution of the Fokker-Planck equation
\begin{equation} \label{eq:F-P}
\pderiv{P} \ = \ \left(\Lop_0 + \Lop_{coll}\right) P \ .
\end{equation}
The  first  term  describing  the  evolution  of the  system,  as  defined  by
\eref{eq:eqs-motion}, can be written as
\begin{equation} \label{eq:L0}
\Lop_0 P = \sum_{i,j} \left(\m{A}_{ij} \frac{\partial x_j P}{\partial x_i} +
\frac{\m{D}_{ij}}{2} \frac{\partial^2 P}{\partial x_i \partial x_j}\right) \ , 
\end{equation}
where  the $2N$  vector $\m{x}  = (q_1,  q_2, \ldots,  q_N, p_1,  p_2, \ldots,
p_N)$, and the $2N \times 2N$ matrices $\m{A}$ and $\m{D}$ are
\begin{equation} \label{eq:mat-0}
\m{A} = \left(
\begin{array}{cc}
\m{0} & -\m{1} \\
\omega^2\m{G} & \lambda\m{R}
\end{array}
\right) \  ; \ \quad 
\m{D} = \left(
\begin{array}{cc}
\m{0} & \m{0} \\
\m{0} & 2\lambda k_B T(\m{R} + \eta\m{S})
\end{array}
\right) \,
\end{equation}
with $\m{0}$ and $\m{1}$ the null and unit $N \times N$ matrices respectively,
\begin{equation} \label{eq:mat-1}
\m{R}_{ij} \ = \ \delta_{i,j}\left(\delta_{i,1} + \delta_{i,N}\right) \ ,
\quad \ \m{S}_{ij} \ = \ \delta_{i,j}\left(\delta_{i,1} - \delta_{i,N}\right)
\ , 
\end{equation}
and $\m{G}$ is the negative of the Laplacian,
\begin{equation} \label{eq:mat-2}
\m{G}_{ij} \ = \ 2\delta_{i,j} - \delta_{i+1,j} - \delta_{i,j+1} \ .
\end{equation}
Moreover, we introduce the normalized bath temperatures difference 
$\eta=\Delta T/T=(T_+-T_-)/T$.
Finally, the second term in \eref{eq:F-P} associated to 
stochastic collisions reads
\begin{equation} \label{eq:Lcoll}
\Lop_{coll} P = \gamma \sum_{j=1}^{N-1} \left[
P(\ldots ,p_{j+1},p_j, \ldots) - P(\ldots ,p_j, p_{j+1}, \ldots) \ \right].
\end{equation}
Each term in the sum expresses the probability balance for each 
elementary process in which momenta of the pair $j,j+1$ are 
exchanged.

As  we mentioned above,  this type  of dynamics  with conservative  noise, was
originally  introduced in  \cite{BBO06} where,  however, only  the equilibrium
case was studied. Here we consider the nonequilibrium situation. Moreover, the
collision  term \eref{eq:Lcoll}  we consider  here has  two  main differences:
first, in the present case, only collisions of pairs (instead of triplets) are
necessary.  Second,  in \cite{BO05}, each  evolution step is  an infinitesimal
variation of  the momenta onto the constant-energy  hypersurface.  This allows
to define a generator of the  process as a differential operator acting on the
$\vec{p}$-space.  On  the contrary,  in the present  case the  process remains
intrinsically discontinuous.

\section{Covariance Matrix}
\label{sec:cov}

Consider the covariance matrix $\mC$ for the dynamics  \eref{eq:eqs-motion},
which we write as
\begin{equation} \label{eq:covmat}
\m{C} = \left(
\begin{array}{cc}
  {\bf U}       &   {\bf Z}\\
  {\bf Z}^\dag  &   {\bf V}
\end{array}
\right) \ ,
\end{equation}
where,
\begin{equation} \label{defcov}
\mU_{i,j} = \cor{q_i q_j} \ ,  \quad 
\mV_{i,j} = \cor{p_i p_j} \ ,  \quad 
\mZ_{i,j} = \cor{q_i p_j} \ ,
\end{equation}
are three  $N\!\times\! N$ matrices,  the square brackets $\langle  . \rangle$
denote  an  average  over   $P(\vec{q},\vec{p},t)$,  and  $\dag$  denotes  the
transpose operation. There  is no need to include  mean values, since $\langle
p_i \rangle = \langle q_i \rangle = 0$.

Note that the matrices $\mU$ and $\mV$ are symmetric by definition. 
The evolution equation for $\mC$ can be written
\begin{equation} \label{eq:cdot}
\dot {\bf C} \; = \; \dot {\bf C}_0+\dot {\bf C}_{coll} \ ,
\end{equation} 
where (see {\em e.g.} equation (63) in \cite{LLP03}),
\begin{equation} \label{eq:cov0}
\dot {\bf C}_0 = {\bf D} - {\bf A }{\bf C} - {\bf C}{\bf A}^\dag \ .
\end{equation} 
The collision term 
$\dot{\bf C}_{coll}$ is evaluated upon
multiplying \eref{eq:Lcoll} by $x_ix_j$ and 
thereby integrating over phase space. We obtain 
\begin{equation}
\dot {\bf C}_{coll} = -\gamma
\left(
\begin{array}{cc}
 {\bf 0}              &   {\bf Z} {\bf G} \\
 {\bf G} {\bf Z}^\dag  &     {\bf W}
\end{array}
\right)
\end{equation}
where the auxiliary $N\!\times \!N$ matrix ${\bf W}$ is defined by
\begin{equation}
\mW_{ij} \equiv
\cases{\mV_{i-1,j-1} + \mV_{i+1,j+1} - 2\mV_{i,j} & $i = j$
\\ \mV_{i-1,j} + \mV_{i,j+1} - 2\mV_{i,j} & $i - j = -1$
\\ \mV_{i+1,j} + \mV_{i,j-1} - 2\mV_{i,j} & $i - j = 1$
\\ \mV_{i+1,j} + \mV_{i-1,j} + \mV_{i,j-1} + \mV_{i,j+1}
- 4\mV_{i,j} & $|i - j| > 1$
}
\label{wmat}
\end{equation}

Equation \eref{eq:cdot} is thus exact and closed and describes the approach to
the nonequilibrium  steady state. In  the present work  we aim at  finding its
stationary  solution, which  amounts to  solving the  set of  linear equations
\numparts
\begin{eqnarray}\label{eq:statsol}
 \mZ^\dag = -\mZ \ 
\label{eq:statsol-a}  ,\\ 
 \mV = \omega^2\mU\mG + \lambda\mZ\mR + \gamma\mZ\mG \ ,
\label{eq:statsol-b} \\
  \omega^2\left(\mG\mZ+\mZ^\dag\mG\right) + \lambda\left(\mR\mV + \mV\mR\right)
+ \gamma\mW = 2\lambda\kB T\left(\mR+\eta\mS\right) \ . 
\label{eq:statsol-c}
\end{eqnarray}
\endnumparts
Note that for $T^+=T^-=T$,  namely $\eta=0$,  these  equations 
admit the equilibrium solution
\begin{equation} \label{eq:equil}
\mU_\mathrm{eq} = \frac{\kB T}{\omega^2}\mG^{-1} \ , \quad \mV_\mathrm{eq} =
\kB T \m{1} \ , \quad \mZ_\mathrm{eq} = \m{0} \ .
\end{equation}
For  $\eta \ne  0$,  analogously to  what  found in  purely stochastic  models
\cite{GKR07,LY07},  we  expect  the  onset  of  a non-zero  heat  flux  to  be
accompanied by the appearance of non-diagonal terms.

In    the    next    Section     we    solve    analytically    the    problem
(\ref{eq:statsol-a}-\ref{eq:statsol-c})  by  means  of  a  suitable  continuum
approximation.   The  idea  is  to  replace  the  finite-difference  equations
(\ref{eq:statsol-a}-\ref{eq:statsol-c})  with a  set  of partial  differential
equations.  Before entering  the technical  details, it  is useful  to briefly
anticipate  the  main  outcomes  of  our calculation.  The  temperature  field
$T_i=\langle  p^2_i\rangle$ along  the  chain,  as a  function  of the  scaled
variable $y\equiv 2i/N -1$ can be expressed as
\begin{equation} \label{eq:T1}
T(y) \;=\; T + \Delta T \, \Theta(y) \ ,
\end{equation}
where $\Theta(y)$ is defined through the following series,
\begin{equation} \label{eq:T2}
\Theta(y) \;=\; \frac{\sqrt{2}}
  {(\sqrt{8}-1)\zeta(3/2)}\sum_{\mathrm{odd} \ n} n^{-3/2}
    \cos\left(\frac{n\pi}{2}(y+1)\right) \ ,
\end{equation}
where $\zeta(3/2) = 2.612375348 \ldots  $ is the Riemann $\zeta$-function.  It
can be seen  that $\Theta(y)$ is an odd function of  $y$ such that $\Theta(\pm
1)=\mp 1/2$.  The leading term of the stationary energy current (see below for
the exact definition) is
\begin{equation} \label{J:8}
J \;=\; \frac{\mathcal{J}}{\sqrt{N}} \;=\; \frac{\Delta T}{8(\sqrt{8}-1)\zeta(3/2)} \,
\sqrt{\frac{\pi^3\omega^3}{\gamma N}}\ 
\end{equation}
As a consequence, the effective conductivity is
\begin{equation} \label{conduct}
\kappa \equiv \frac{J}{\Delta T/N} \; = \;
 \frac{1}{8(\sqrt{8}-1)\zeta(3/2)} \, \sqrt{\frac{\pi^3\omega^3 N}{\gamma}}\ 
\end{equation} 
Comments on the  physical meaning of these formulae will be  given in the last
Section.

\section{Analytical solution}
\label{sec:perturb}

The  solution of  (\ref{eq:statsol-a}-\ref{eq:statsol-c})  can be  efficiently
determined numerically by exploiting  the sparsity of the corresponding linear
problem, as  well as the symmetries of  the unknowns ${\bf U}$,  ${\bf V}$ and
${\bf Z}$. This  approach has been followed in  \cite{DLLP08}.  Here, we solve
the problem  analytically treating the  ``lattice" equations in  the continuum
approximation. It  must be  first recognized that  the correct scaling  is not
known  \textit{a priori},  but rather  inferred from  the  numerical solution.
Therefore,  the  correctness  of  the  results has  to  be  checked  \textit{a
posteriori} by consistency.

\subsection{The continuum limit}

\begin{figure}[!t]
\begin{center}
  \includegraphics[scale=1.2]{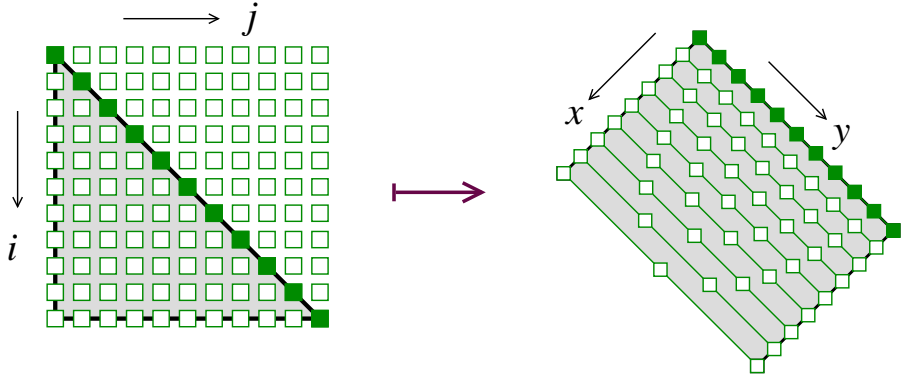}
  \caption{Schematic  representation of  the mapping  from the  matrix indexes
  $i,j$ (left) to  the continuous variables $(x,y)$ (right).   The latter vary
  in the  domain $\mathcal{D}$, definition \eref{def:D}  (shaded region).  The
  square  symbols  represent  matrix  elements, (bolded  along  the  diagonal,
  $i=j$).   Diagonals  are  parametrically  obtained  as  $x=$  constant,  and
  antidiagonals as  $y=$ constant.  The  denominator in the definition  of $y$
  \eref{def:xy}, takes into account that the length of the diagonals depend on
  their value of $x$ so that, the domain of $y$ is independent of $x$.
\label{fig:xy}}
\end{center}
\end{figure}

The first step consists in mapping the discrete variables $i$ and $j$ into two
suitable continuous variables  $x$ and $y$, so that  an $N\!\times\! N$ matrix
${\bf M}_{ij}$ can be transformed  into a ``field variable" ${\bf M}(x,y)$ and
the  associated   discrete  equation   turned  into  a   partial  differential
equation. In  order to do so, it  is first necessary to  introduce a smallness
parameter that vanishes as $N \to \infty$. In \cite{DLLP08}, it was found that
while neighbouring  elements along the  diagonal differ by  $\Or(1/N)$, across
the diagonal  the difference is $\Or(1/\sqrt{N})$. This  suggests defining the
smallness  parameter as  $\e=1/\sqrt{N}$.  In addition,  it  is convenient  to
introduce  a further  stretching of  the longitudinal  variable $y$  so  as to
ensure a constant  elongation in the $(x,y)$ representation.  This is achieved
through the following transformation
\begin{equation} \label{def:xy}
x \equiv (i-j)\e \ ; \qquad y \equiv \frac{(i+j)\e^2 -1}{1-|i-j|\e^2}\ .
\end{equation}
that is  schematically represented also in figure  \ref{fig:xy}. The nonlinear
transformation complicates the  expansions along $y$, but is  essential to set
the  boundary conditions  correctly.  Although \eref{def:xy}  is singular  for
$|i-j|=N$, this  is harmless, since its  location diverges to  infinity and is
thus located outside the region of interest. In the infinite volume limit, the
variables $(x,y)$ belong in the domain
\begin{equation} \label{def:D}
\mathcal{D} \equiv \left\{(x,y) | x\in[0,\infty); \ y\in[-1,1] \right\}
\end{equation}
Note that  $x=const$ corresponds to  moving along a diagonal  direction, $x=0$
corresponding to the main diagonal.

In   order    to   determine   the   continuum   limit    of   the   equations
(\ref{eq:statsol-a}-\ref{eq:statsol-c}),  it   is  necessary  deal   with  the
infinitesimal changes of $x$ and $y$ that arise from $\Delta i$ and $\Delta j$
shifts of  $i$ and $j$.  It  is convenient to introduce  the integer functions
$f,s:\mathbb{Z}^2\mapsto\mathbb{Z}$
\begin{equation} \label{eq:fs}
f(\Delta i,\Delta j) \equiv \Delta i - \Delta j \ , \qquad 
s(\Delta i,\Delta j) \equiv \Delta i + \Delta j \ .
\end{equation}
With the help of these shift  functions, the coordinates of a point shifted by
$(\Delta i,\Delta j )$ reads
\begin{equation} \label{eq:xyderiv}
x' = x + f\e \ ; \qquad  y' = \frac{(i+j)\e^2-1+s\e^2}{1-(i-j)\e^2 -f\e^2}
\end{equation}
where we assume that $i\ge j$ to get rid of the absolute value. Accordingly,
\begin{equation}
y' = \left(y + \frac{\e^2s}{1-\e x}\right)
\frac{1}{1 - \e^2f/(1-\e x)} \ ,
\end{equation}
and, up to fourth order in $\e$,
\begin{equation}
y' = \left(y + \e^2 s(1+\e x +\e^2 x^2)\right) \left(1+\e^2 f(1+\e x+\e^2x^2)
+\e^4 f^2 \right) \ ,
\end{equation}
which is conveniently written as
\begin{equation} \label{eq:yderiv}
y' = y + \e^2 \left(1+\e x +\e^2 (x^2+f)\right)(s+fy) \equiv y + \e^2 R_{f,s}
\ , 
\end{equation}
where
\begin{equation} \label{def:R}
R_{f,s} = \left[1+\e x +\e^2 (x^2+f)\right](fy+s) \ .
\end{equation}
With  these definitions,  an infinitesimal  change  in $x$  involves terms  of
$\Or(\e)$ and an  infinitesimal change in $y$ generates  terms of $\Or(\e^2)$,
$\Or(\e^3)$  and  $\Or(\e^4)$.   However,  for  the estimate  of  the  leading
contributions,   it  is   sufficient   to  consider   $R_{f,s}  =   \left(1+\e
x\right)(fy+s)$.

Altogether, the  above relations provide  a useful tool for  investigating the
continuum limit. For  later applications, the above results  are summarized in
the rule,
\begin{equation}
{\bf M}_{i+\Delta i,j+\Delta j} = {\bf M}(x+f\e, y+\e^2R_{f,s}) .
\label{rule}
\end{equation}
that is written in a convenient form  for an expansion in powers of $\e$. Here
and  in  what follows,  we  keep the  bold-face  notation  for the  continuous
functions derived from the matrix variables.

\subsection{Field variables}

The disadvantage of representation (\ref{defcov}) is that $\mU_\mathrm{eq}$ is
a  full  matrix  whose  diagonal  elements  are  $\Or(N)$.  This  hinders  the
formulation  of a proper  perturbation scheme  to compute  the non-equilibrium
corrections.   For  the  sake  of   the  numerical  solution  carried  out  in
\cite{DLLP08},  this difficulty has  been overcome  by looking  at correlators
involving  relative  rather than  absolute  displacements, i.e.,  $\mZ'_{i,j}=
\langle(q_{i}-q_{i+1})p_j\rangle$               and               $\mU'_{i,j}=
\langle(q_{i+1}-q_{i})(q_{j+1}-q_{j})\rangle$.      In    fact,     in    this
representation, $\mU'$ turns  out to be diagonal at  equilibrium with diagonal
elements of $\Or(1)$.  On the  other hand, $\mZ'_{i,j}$ loses the antisymmetry
of    $\mZ_{i,j}$,   a    very    useful   property    for   our    analytical
treatment. Therefore,  we have decided to  keep the definition of  $\mZ$ as in
(\ref{defcov})  while  introducing a  new  matrix  $\mY_{i,j} \equiv  \omega^2
\langle(q_{i+1}-q_{i})(q_{j+1}-q_{j})\rangle$, which is conveniently expressed
in terms of $\mU$ as,
\begin{equation} \label{def:Y}
\mY_{i,j} \equiv  
\omega^2[\mU_{i,j} - \mU_{i,j+1} - \mU_{i+1,j}  + \mU_{i+1,j+1}] \ .
\end{equation} 
Note that the diagonal elements are proportional to the average bond potential
energy $\Phi_i$,
\begin{equation} \label{def:Ydiag}
\mY_{i,i} \equiv  
\omega^2 \langle (q_{i+1}-q_i)^2 \rangle \equiv 2\Phi_i\ .
\end{equation}

The next step consists in choosing  the proper order of magnitude of the three
fields $\mV$, $\mY$, and $\mZ$. This  will be done by exploiting the knowledge
of the  equilibrium case and  the information arising from  previous numerical
solution \cite{DLLP08}.  First of all,  since $\mY_{i,i}$ and  $\mV_{i,i}$ are
proportional to the mean potential  and kinetic energy, respectively, they are
both  of $\Or(1)$  as  in equilibrium.  On  the other  hand, the  off-diagonal
elements turn out to be of  $\Or(\e)$. Hence, for consistency of the continuum
approximation, we must consider independently diagonal and off-diagonal (bulk)
entries of $\mV$ and $\mY$.  The matrix $\mZ$ exhibits a somehow complementary
behaviour. Since it is antisymmetric in $x$, there are no diagonal terms,
\begin{equation} \label{eq:ssa-0}
\mZ(0,y) = 0 \ ,
\end{equation}
while the numerics  suggests that in the bulk it is  $\Or(1)$.  We thus define
the following field variables: in the bulk ($i\ne j$, $x>0$)
\begin{equation} 
 \label{eq:exp-b} \fl
 \mY_{i,j}  \ = \   \e\mY(x,y) + \hot \ ,\\
 \mV_{i,j}  \ = \   \e\mV(x,y) + \hot \ ,\\
 \mZ_{i,j}  \ = \     \mZ(x,y) + \hot \ .
\end{equation}
and for the diagonal ($i=j$, $x=0$) terms
\begin{equation} \label{eq:exp-d}
 {\mV}_{i,i}  \ = \  T(y) + \hot \ ,\qquad
 {\mY}_{i,i}  \ = \  2\Phi(y) + \hot \ ,
\end{equation}
The scaling properties  of the first corrections to the  leading order are not
known and  the comparison  with the numerical  results discussed in  the final
section show the existence of a singular dependence on $\e$. As a consequence,
it is not  possible to set up a standard perturbation  expansion scheme and it
is  therefore  necessary to  rely  on  expressions  dominated by  the  leading
contributions.  In  the  next  section  we demonstrate  that  by  manipulating
equations    (\ref{eq:statsol-a},\ref{eq:statsol-b},\ref{eq:statsol-c})    and
suitable  combinations of  them, it  is possible  to obtain  a set  of partial
differential equations whose solution gives the covariance matrices at leading
order.

\subsection{Stationary equation in the bulk}
\label{sec:sol}

Using the  definitions \eref{eq:mat-0},  \eref{eq:mat-1}  and  \eref{eq:mat-2}, the
equation \eref{eq:statsol-b} writes
\begin{equation} \label{eq:ssb-0}
\omega^2\left(2\mU_{i,j}-\mU_{i,j+1}-\mU_{i,j-1}\right) -\mV_{i,j} +
\gamma\left(2\mZ_{i,j}-\mZ_{i,j-1}-\mZ_{i,j+1}\right) = 0 \ .
\end{equation}
The   reader    should   note   that   the   term    $\mZ\mR$   appearing   in
\eref{eq:statsol-b}, can be written as
\begin{equation}
\lambda\mZ\mR =\lambda\left( \mZ(-x,-1) + \mZ(x,1)\right) \ .
\end{equation}
This term  does only  contribute at the  boundaries and consequently,  we have
omitted  it in \eref{eq:ssb-0}.   In the  subsequent treatment,  this omission
will be justified later when we fix the boundary conditions of $\mZ$.

In order  to write the  equations in  terms of the  new variable $\mY$  let us
first rewrite \eref{eq:statsol-b} with $i$ replaced by $i+1$:
\begin{equation} \label{eq:ssb-1}
\fl\omega^2\left(2\mU_{i+1,j}-\mU_{i+1,j+1}-\mU_{i+1,j-1}\right) -\mV_{i+1,j} +
\gamma\left(2\mZ_{i+1,j}-\mZ_{i+1,j-1}-\mZ_{i+1,j+1}\right) = 0 \ .
\end{equation}
Subtracting \eref{eq:ssb-1} from \eref{eq:ssb-0},  and using the definition of
the matrix $\mY$ we obtain
\begin{equation} \label{eq:ssb-discrete}
\begin{array}{l}
\fl \mY_{i,j} - \mY_{i,j-1} + \mV_{i+1,j}  - \mV_{i,j}
+\gamma\left[-2\mZ_{i+1,j}+2\mZ_{i,j} + \mZ_{i+1,j-1}-\mZ_{i,j-1} +\right. \\
\left.  \mZ_{i+1,j+1}-\mZ_{i,j+1}\right]
= 0 \ . 
\end{array}
\end{equation}
With   the   help   of   rule   \eref{rule},   the   continuous   version   of
\eref{eq:ssb-discrete} in the bulk is readily written as
\begin{equation} \label{eq:ssb-2}
  \begin{array}{l}
\fl\mY(x,y) - \mY(x+\e,y+\e^2R_{1,-1}) +
  \mV(x+\e,y+\e^2R_{1,1})  - \mV(x,y) + \\
\fl \gamma\left[-2\mZ(x+\e,y+\e^2R_{1,1})+  2\mZ(x,y)+
  \mZ(x+2\e,y+\e^2R_{2,0}) - \right. \\
\fl \left. \mZ(x+\e,y+\e^2R_{1,-1})+\mZ(x,y+\e^2R_{0,2})-
\mZ(x-\e,y+\e^2R_{-1,1})\right]  = 0 \ .
\end{array}
\end{equation}
The leading order of \eref{eq:ssb-2} is of $\Or(\e^2)$ and can be written as
\begin{equation} \label{eq:ssb-3}
\mathbf\Omega_x(x,y) = 0 \, ,
\end{equation}
where  the  subscripts denote  the  partial  derivative  with respect  to  the
corresponding  variable and  for reasons  that will  be clear  below,  we have
introduced the function
\begin{equation} \label{def:Omega}
\mathbf\Omega(x,y)\equiv\mY(x,y) - \mV(x,y) \ .
\end{equation}
Furthermore, by exchanging $i$ with $j$ in equation \eref{eq:ssb-discrete} and
adding the result to \eref{eq:ssb-discrete}, we find a symmetrized equation in
the bulk, given by
\begin{equation} \label{eq:ssb-4}
\begin{array}{l}
\fl 2\mY_{i,j} - \mY_{i,j-1} - \mY_{i-1,j} + \mV_{i+1,j} +
\mV_{i,j+1}     -    2\mV_{i,j}     
+\gamma\left[\mZ_{i,j+1} -\mZ_{i+1,j}    +
 \right.\\
\left. \mZ_{i+1,j-1}-\mZ_{i-1,j+1}+\mZ_{i-1,j}-\mZ_{i,j-1}\right] = 0 \ .
\end{array}
\end{equation}
The  continuous  version of  \eref{eq:ssb-4} is
\begin{equation} \label{eq:ssb-5}
\begin{array}{l}
\fl2\mY(x,y) - \mY(x+\e,y+\e^2 R_{1,-1}) - \mY(x-\e,y+\e^2
R_{-1,-1})  -2\mV(x,y) + \\
\fl \mV(x+\e,y+\e^2 R_{1,1}) + \mV(x-\e,y+\e^2R_{-1,1}) +
 \gamma\left[\mZ(x-\e,y+\e^2 R_{-1,1}) - \right. \\
\fl \mZ(x+\e,y+\e^2 R_{1,1}) + \mZ(x+2\e,y+\e^2 R_{2,0})-\mZ(x-2\e,y+\e^2
R_{-2,0}) + \\
\fl \left. \mZ(x-\e,y+\e^2 R_{-1,-1})-\mZ(x+\e,y+\e^2 R_{1,-1}) \right] = 0 \ ,
\end{array}
\end{equation}
whose leading contribution yields
\begin{equation} \label{eq:ssb-6}
  -\mathbf\Omega_{xx}(x,y)  +   2\left[\mY_y(x,y)  +  \mV_y(x,y)\right]  +
2\gamma\mZ_{xxx}(x,y) = 0 \ .
\end{equation}
By using \eref{eq:ssb-3}, the above equation becomes
\begin{equation} \label{eq:sol-0}
\mY_y(x,y)  +  \mV_y(x,y)  + \gamma\mZ_{xxx}(x,y) = 0 \ .
\end{equation}
Furthermore,   integrating    \eref{eq:ssb-3}   on   $x$    we   obtain   that
$\mathbf\Omega(x,y)$ does not depend on the transversal coordinate $x$, namely
\begin{equation} \label{eq:sol-1}
\mathbf\Omega(x,y) \equiv \mathcal{F}(y) \ .
\end{equation}
By using this in \eref{eq:sol-0} to replace $\mY$ with $\mV$, we
obtain
\begin{equation} \label{eq:sol-2}
\mV_y(x,y) = -\frac{\gamma}{2}\mZ_{xxx}(x,y) - \frac{1}{2}\mathcal{F}(y) \ .
\end{equation}

Proceeding as  before, with the  help of \eref{wmat}, the  stationary equation
\eref{eq:statsol-c} in the continuum is
\begin{equation}
\begin{array}{l} \label{eq:ssc-0}
\fl\omega^2\left[\mZ(x+\e,y+\e^2R_{1,-1}) + \mZ(x-\e,y+\e^2R_{-1,1}) -
\mZ(x-\e,y+\e^2R_{-1,-1}) -\right. \\
\fl\left. \mZ(x+\e,y+\e^2R_{1,1})\right] +  \gamma\left[\mV(x+\e,y+\e^2R_{1,1}) +
 \mV(x-\e,y+\e^2R_{-1,-1}) + \right. \\
\fl\left. \mV(x+\e,y+\e^2R_{1,-1}) + \mV(x-\e,y+\e^2R_{-1,1}) -
4\mV(x,y)\right] = 0 \ .
\end{array}
\end{equation}
The leading order contribution is of $\Or(\e^3)$,
\begin{equation} \label{eq:ssc-1}
\mV_{xx}(x,y) =  \frac{2\omega^2}{\gamma}\mZ_{xy}(x,y) \ .
\end{equation}
By integrating in $x$, we obtain
\begin{equation} \label{eq:sol-3}
\mV_{x}(x,y) =  \frac{2\omega^2}{\gamma}\mZ_{y}(x,y) + {\mathcal G}(y) \ ,
\end{equation}
where  ${\mathcal G}(y)$  is  a  suitable integration  constant  that will  be
determined by imposing the boundary conditions.  Now, taking the derivative of
\eref{eq:sol-2} \wrt $x$, and the  derivative of \eref{eq:sol-3} \wrt $y$, and
summing the  results, we obtain a  differential equation for  the behaviour of
$\mZ$ in the bulk,
\begin{equation} \label{sol:4}
\mZ_{xxxx}(x,y) - \frac{4\omega^2}{\gamma^2}\mZ_{yy}(x,y) = \frac{2}{\gamma}
{\mathcal G}_y(y) \ .
\end{equation}
This  is the  general  equation, whose  solution  yields the  behavior of  the
various fields in the bulk.

\subsection{Boundary conditions}

In this  section we  impose all boundary  conditions. The  various constraints
allow not  only to uniquely determine the  behaviour in the bulk,  but also to
establish  a  link with  the  physically  relevant  observables, such  as  the
temperature profile. Analogously to the  previous section, we proceed into two
steps by  separately analysing the implications of  \eref{eq:statsol-b} and of
\eref{eq:statsol-c}.

By setting $i=j$ in \eref{eq:ssb-discrete}, we obtain
\begin{equation} \label{eq:ssb-7}
\begin{array}{l}
\fl \mY_{i,i} - \mY_{i,i-1} + \mV_{i+1,i}  - \mV_{i,i}
+\gamma\left[-2\mZ_{i+1,i}+2\mZ_{i,i}+\mZ_{i+1,i-1}-\mZ_{i,i-1}+\right.\\
\left. \mZ_{i+1,i+1}-\mZ_{i,i+1}\right] = 0 \ . 
\end{array}
\end{equation}
We recall that in order to avoid the complication of the absolute value in the
denominator of  \eref{eq:yderiv}, we have assumed that  $i\ge j$. Accordingly,
the  use  of  \eref{eq:yderiv}  requires  to consider  always  the  lower  (by
convention) triangle of all the  matrices. In order to satisfy this condition,
we exploit the antisymmetry of $\mZ$ to obtain
\begin{equation*} 
\mY_{i,i} - \mY_{i,i-1} + \mV_{i+1,i}  - \mV_{i,i}
+\gamma\left(-\mZ_{i+1,i}+\mZ_{i+1,i-1}-\mZ_{i,i-1}\right)
= 0 \ ,
\end{equation*}
and, in field variables, 
\begin{equation} \label{eq:ssb-8}
\begin{array}{l}
\fl 2\Phi(y)   -  \mY(\e,y+\e^2R_{1,-1})  +   \mV(\e,y+\e^2R_{1,1})  -
T(y) + \\
\gamma\left(-\mZ(\e,y+\e^2R_{1,1})
  +\mZ(2\e,y+\e^2R_{2,0})-\mZ(\e,y+\e^2R_{1,-1})\right) = 0 \ . 
\end{array}
\end{equation}
The leading contribution  of \eref{eq:ssb-8} is $\Or(1)$, as  expected for the
diagonal terms, yielding a boundary condition for $\mathbf\Omega$:
\begin{equation} \label{eq:ssb-9}
\Omega(y) = 2\Phi(y) - T(y) = 0 \ .
\end{equation}
This last  expression is just  a local version  of the virial theorem  for the
harmonic    oscillators.

The reader  can verify that the  leading term of \eref{eq:ssb-2}  in the upper
diagonal  ($i=j-1$) does not  give further  information.  However,  adding the
equation  for the  upper  diagonal  to \eref{eq:ssb-7},  we  obtain, in  field
variables,
\begin{equation} \label{eq:ssb-10}
\begin{array}{l}
\fl \mY(x+\e,y+\e^2R_{1,1})-\mY(x+\e,y+\e^2R_{1,-1}) + 
T(y+\e^2R_{0,2}) - T(y) + \\
\fl \gamma\left[-2\mZ(x+\e,y+\e^2R_{1,1})
  -\mZ(x+\e,y+\e^2R_{1,3})+\mZ(x+2\e,y+\e^2R_{2,2})+ \right. \\
\fl \left. \mZ(x+2\e,y+\e^2R_{2,0}) - \mZ(x+\e,y+\e^2R_{1,-1})\right] = 0 \ .
\end{array}
\end{equation}
This  equation gives  rise to  two relations  of leading  order. The  first is
redundant as  it confirms that  $\mZ$ is zero  along the diagonal.  The second
relation is, instead, a differential equation for $T(y)$,
\begin{equation} \label{eq:ssb-11}
T_{y}(y) + \gamma \mZ_{xx}(0,y) = 0 \ . 
\end{equation}
It  allows  determining the  temperature  profile,  once  $\mZ(x,y)$ has  been
determined.

We now turn out attention  to \eref{eq:statsol-c}. Along the diagonal ($i=j$),
it is  \footnote{The reader  can easily check  that if one  identifies $T(-1)$
with the left temperature $T^+$  and $T(+1)$ with the right temperature $T^-$,
then the  boundary terms in  \eref{eq:statsol-c} cancel to each  other, namely
$\left(\mR\mV + \mV\mR\right) = 2\kB T\left(\mR+\eta\mS\right)$.}
\begin{equation}\label{eq:ssc-2a}
\gamma(2\mV_{i,i}-\mV_{i-1,i-1}-\mV_{i+1,i+1}) + 2\omega^2(\mZ_{i,i-1}-
\mZ_{i+1,i}) = 0  \ .
\end{equation}
It is straightforward to show that the above equation is equivalent to
\begin{eqnarray}\label{eq:ssc-2b}
\frac{\gamma}{2}(\mV_{i,i}-\mV_{i-1,i-1}) + \omega^2 \mZ_{i,i-1} = -J
\end{eqnarray}
where the integration  constant $J$ is nothing but the  average heat flux.  In
fact, the energy flux $J_i$ between the  particles $i-1$ and $i$ is the sum of
two contributions,  a deterministic one  $\Jd_i$, due to the  interaction with
the neighbours, and a stochastic one $\Js_i$, originating from the collisions,
\begin{equation} \label{J:1}
J_i \;  = \; \Jd_i   + \Js_i  \,
\end{equation}  
with
\begin{eqnarray} \label{J:2}
 \Jd_i &\;\equiv\; & \omega^2 \langle q_{i-1} p_i \rangle 
 \; = \; \omega^2 \mZ_{i-1,i} \ ,\\
 \Js_i &\; \equiv\; & \frac{\gamma}{2} \left(\langle p^2_{i-1}\rangle - \langle p^2_{i}\rangle \right)
\; =\;  \frac{\gamma}{2} \left( \mV_{i-1,i-1} - \mV_{i,i}\right) \ ,
\end{eqnarray}
where in both definitions we have  adopted the convention that a positive flux
corresponds  to energy  travelling  from smaller  to  larger $i$  coordinates.
Accordingly,  \eref{eq:ssc-2b} states the  well known  physical fact  that the
heat  flux is  constant along  the chain  (i.e., independent  of $i$).  In the
continuum limit, equation \eref{eq:ssc-2b} writes
\begin{equation} \label{eq:ssc-2c}
\frac{\gamma}{2}\left[T(y)-T(y+\e^2R_{0,-2})\right] + 
\omega^2 \mZ(\e,y+\e^2R_{1,-1}) = -J \ ,
\end{equation}
The leading contribution of the l.h.s. is  of $\Or(\e)$ and so must be $J$ ($J
= \mathcal{J}\e$). As a result, we can write,
\begin{equation} \label{eq:ssc-3}
\omega^2 \mZ_x(0,y) = -\mathcal{J} \ ,
\end{equation}
This  is a  relevant  piece of  information  that will  allow  us to  uniquely
determine $\mZ(x,y)$ in the bulk.

Finally, for the upper diagonal ($i=j+1$), \eref{eq:statsol-c} becomes
\begin{equation} \label{eq:ssc-4}
\begin{array}{l}
\fl \omega^2\left[\mZ(0,y) - \mZ(2\e,y+\e^2R_{2,2}) + \mZ(2\e,y+\e^2R_{2,0}) -
\mZ(0,y+\e^2R_{0,2})\right] + \\
 \gamma\left[\mV(2\e,y+\e^2R_{2,0}) + \mV(2\e,y+\e^2R_{2,2}) 
 - 2\mV(\e,y+\e^2R_{1,1})\right] = 0 \ ,
\end{array}
\end{equation}
from where we obtain to leading order
\begin{equation} \label{eq:ssc-5}
2\omega^2\mZ_y(0,y) + \gamma\mV_{x}(0,y) = 0 \ ,
\end{equation}
that, by virtue of \eref{eq:ssa-0}, implies
\begin{equation} \label{eq:ssc-6}
\mV_{x}(0,y) = 0  \ .
\end{equation}
For  \eref{eq:statsol-c},  combinations of  the  diagonal  and upper  diagonal
relations give no further information.

\subsection{Solution of the equations}
\label{sec:obs}

In this section we solve the differential equations of covariance matrices, to
leading order  in $\e$.  From  this solution we derive  analytical expressions
for the temperature  profile and the energy flux.  We  start noticing that the
function ${\mathcal  G}(y)$ appearing in \eref{eq:sol-3}  is identically equal
to  zero.   This  is seen  by  setting  $x=0$  and using  \eref{eq:ssa-0}  and
\eref{eq:ssc-6}. As a result, \eref{sol:4} simplifies to,
\begin{equation} \label{sol:5}
\mZ_{xxxx}(x,y) - \frac{4\omega^2}{\gamma^2}\mZ_{yy}(x,y) = 0 \ .
\end{equation}
The form of the  above equation suggests to look for a  solution by the method
of  separation  of variables.   Furthermore,  the  numerical  solution of  the
stationary  solution  \eref{eq:statsol-a} suggests  that  $\mZ(x,y)=0$ at  the
boundaries of  the domain $\mathcal{D}$.   Therefore, we assume  the following
{\em Ansatz}
\begin{equation} \label{eq:ansatz-1}
\mZ(x,y) = \sum_n B_n(x)
\sin\left[\bn\left(y+1\right)\right] \ , \qquad 
\bn \equiv\frac{n\pi}{2}. 
\end{equation}
which, upon  substitution into \eref{sol:5}, gives
\begin{equation} \label{eq:aux-1}
\frac{\mathrm{d}^4 B_n}{\mathrm{d}x^4} =
-\left(\frac{n\pi\omega}{\gamma}\right)^2 B_n \ .
\end{equation}
This  is  readily  solved  by   finding  the  four  roots  of  the  associated
characteristic polynomial. Two of the  four eigenvalues having a positive real
part  would  lead  to  an  unphysical   divergence  in  $x$  and  have  to  be
discarded.  Another constraint is  imposed, by  recalling that  $\mZ(0,y) =0$.
Altogether, the coefficients $B_n$ can be written as,
\begin{equation} \label{eq:aux-3}
B_n(x) = A_n  \exp(-\an x) \sin(\an x)\ , \qquad
\an \equiv  \left(\frac{n\pi\omega}{2\gamma}\right)^{1/2} \ .
\end{equation}
Finally, the constants $A_n$ can be determined by imposing \eref{eq:ssc-3}
\begin{equation} \label{eq:Z0}
\mZ(x,y) = -\frac{2\mathcal{J}}{\omega^2} \sum_{\mathrm{odd} \ n} 
\frac{1}{\an\bn} \exp(-\an x) \sin(\an x)\sin(\bn(y+1)) \ .
\end{equation}
The  only  remaining unknown,  $\mathcal{J}$,  can  be  finally determined  by
imposing that the temperature profile interpolates between $T_+$ and $T_-$. By
integrating \eref{eq:ssb-11} in $y$, we find
\begin{equation} \label{eq:ssb-T}
T(y) = T - \gamma \int_0^y \mZ_{xx}(0,s) \ \mathrm{d}s \ ,
\end{equation}
where we have identified $T(0) = T = (T^++T^-)/2$.  By substituting expression
\eref{eq:Z0} into \eref{eq:ssb-T} and performing the integral term by term, we
obtain
\begin{equation} \label{T:2}
T(y) = T +\frac{4\gamma \mathcal{J}}{\omega^2} \sum_{\mathrm{odd} \ n}
\frac{\an}{\bn^2}\cos(\bn(y+1)) \ .
\end{equation}
The  value of  $\mathcal{J}$  is obtained  by  imposing $T(-1)  = T^+$.   From
\eref{T:2}, it follows that
\begin{equation} \label{T:4}
\frac{\Delta T}{2} = 8\mathcal{J} \left(\frac{2\gamma}{\pi^3 \omega^3}\right)^{1/2} \sum_{\mathrm{odd} \ n}n^{-3/2} \ .
\end{equation}
Using the formula \cite{GR}
\begin{equation} \label{T:5}
\sum_{\mathrm{odd} \ n}n^{-3/2} =
\frac{\sqrt{8}-1}{\sqrt{8}}\zeta\left(\frac{3}{2}\right) \ ,
\end{equation}
where  the  Riemann  $\zeta$-function  has  been  introduced,  we  obtain  for
$\mathcal{J}$
\begin{equation} \label{T:6}
\mathcal{J} = \left(\frac{\pi^3\omega^3}{\gamma}\right)^{1/2}\frac{\Delta
    T}{8(\sqrt{8}-1)\zeta(3/2)} \ ,
\end{equation}
which  corresponds  to  expression  \eref{J:8}   for  the  heat  flux  in  the
thermodynamic limit. Moreover,  by substituting $\mathcal{J}$ into \eref{T:2},
we obtain  expression \eref{eq:T2} for the temperature  profile.  Finally, the
equation \eref{T:6} allows a unique identification of $\mZ$. From \eref{eq:Z0}
we find
\begin{equation} \label{eq:Z0-a}
\fl \mZ(x,y) = -\frac{\Delta T}{\omega\sqrt{2}(\sqrt{8}-1)\zeta(3/2)}
\sum_{\mathrm{odd} \ n} n^{-3/2} e^{-\an x} \sin(\an x)\sin(\bn(y+1))\ , 
\end{equation}

\section{Discussion and open problems}
\label{sec:concl}

Several  comments are in  order about  the analytical  results derived  in the
previous section, starting from the expression \eref{J:8} for the leading term
of the heat  flux. First of all, we  see that the flux $J$  is proportional to
the temperature difference $\Delta T$. This feature does not only apply to the
leading term, but is a general property which follows from the harmonic nature
of the  underlying dynamics. In more  general contexts, we  expect a nonlinear
response regime to exist. 

Moreover, $J$  is independent of the  strength of the coupling  with the baths
$\lambda$. This can  be physically  understood by  realizing that $\lambda$
plays the r\^ole of the inverse of a contact  resistance. In the thermodynamic
limit, the overall thermal resistance is the sum of the contact plus the bulk  
contribution which eventually dominates, no matter how small is $\lambda$.  
Only for $\lambda=0$, the asymptotic regime cannot be attained (the system
is isolated). The coupling $\lambda$ will presumably manifest itself when
accounting for higher order terms.  

It is interesting  to notice the inverse square root dependence  of $J$ on the
rate $\gamma$ of internal collisions.   The limiting values $\gamma \to 0$ and
$\gamma \to  \infty$, signal a crossover  towards a regime  characterized by a
slower  (faster)  decay of  $J$,  respectively.   This  is the  case,  because
$\gamma=0$ corresponds  to an  integrable dynamics, while  for $\gamma=\infty$
the decay of the heat flux is determined by higher order terms.

As  for  the   temperature  profile,  we  should  stress   that  the  equation
\eref{eq:T2} represents  the first example of an  analytic expression obtained
in  the presence  of anomalous  heat  conduction.  This  is all  the way  more
important,  by recalling that,  as noticed  in \cite{DLLP08},  the temperature
profile of  this stochastic model is quite  similar to that found  in a purely
deterministic system  such as the purely quartic  Fermi-Pasta-Ulam chain. Even
more remarkably, $T(y)$ is a parameter-free function. Indeed, once the profile
is shifted  around the average  temperature and the temperature  difference is
rescaled to unity, the resulting  shape $\Theta(y)$ is not only independent of
$\lambda$  but  also  of  $\gamma$  and  $\omega$.   This  suggests  that  the
temperature  profile  might be  universal  (at least  in  the  limit of  small
temperature  differences  in  truly  nonlinear systems).   Unfortunately  pure
numerics alone is  not sufficient to clarify this issue.   Finally, we wish to
comment on the  singularity observed at the two extrema,  namely for $y$ close
to $-1$ and 1. From \eref{eq:T2}, we find that for $y=-1+\delta y$
\begin{equation} \label{eq:T2b}
\delta \Theta(y) \; \approx \; \sum_{\mathrm{odd} \ n}^{1/\delta y} n^{1/2} \delta y^2
\; \approx \;  \delta y^{1/2} \ ,
\end{equation}
where the  cosine has been approximated with  a parabola and the  sum has been
limited  to $n  < 1/\delta  y$, to  prevent that  the argument  of  the cosine
becomes larger than $\Or(1)$. Altogether  the above equation tells us that the
profile is characterized by a square root singularity.


Although our analysis has allowed us  to determine an exact expression for the
field $\mZ(x,y)$  at leading order in  the bulk, and thus  for the temperature
profile and  the heat current  in the steady  state, the determination  of the
other fields $\mV(x,y)$ and $\mY(x,y)$, requires the knowledge of higher-order
terms. Indeed,  the integration constant ${\mathcal  F}(y)$ in \eref{eq:sol-1}
that helps determining  $\mV(x,y)$ and $\mY(x,y)$ cannot be  obtained from our
analysis.   A  comparison with  numerics  \cite{du}  suggests that  ${\mathcal
F}(y)=0$, but none  of the equations we have analysed  in the previous section
supports  this  observation.   Presumably   one  should  consider  some  other
combinations of the equations (\ref{eq:statsol-a}-\ref{eq:statsol-c}), but the
investigation is  hindered by  the fact that  we are  not entitled to  use any
information on the behaviour of higher-order terms.

\begin{figure}[!t]
\begin{center}
  \includegraphics[scale=0.85]{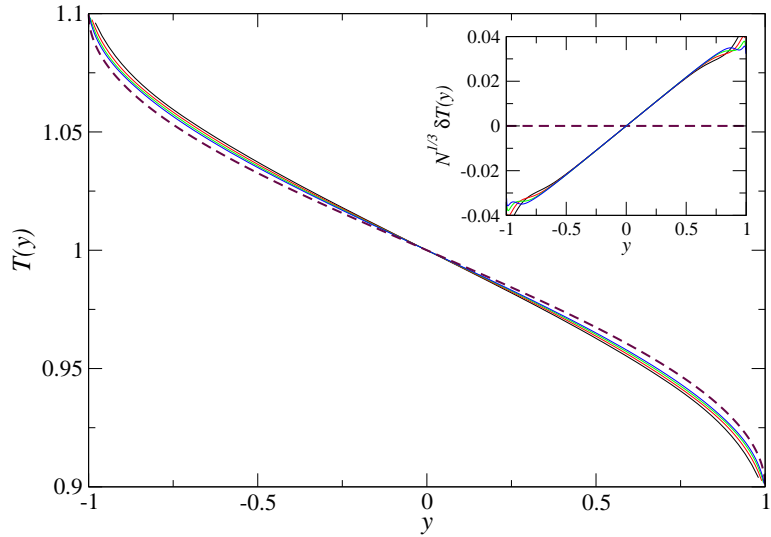}
\caption{Temperature  profile $T(y)$  as  given by  the analytical  expression
  \eref{eq:T1},  for $T_+=1.1$,  $T_-=0.9$,  $\omega=\lambda=\gamma=1$ (dashed
  curve). The solid curves correspond  to the profile $T_{num}$, obtained from
  the numerical  solution of equations (\ref{eq:statsol-a}-\ref{eq:statsol-c})
  for  $N=100,200,400,800$  . The  finite-size  deviations from  \eref{eq:T1},
  $\delta T \equiv T - T_{num}$, rescaled by $N^{1/3}$, are shown in the inset.
\label{fig:T}}
\end{center}
\end{figure}

\begin{figure}[!t]
\begin{center}
  \includegraphics[scale=0.85]{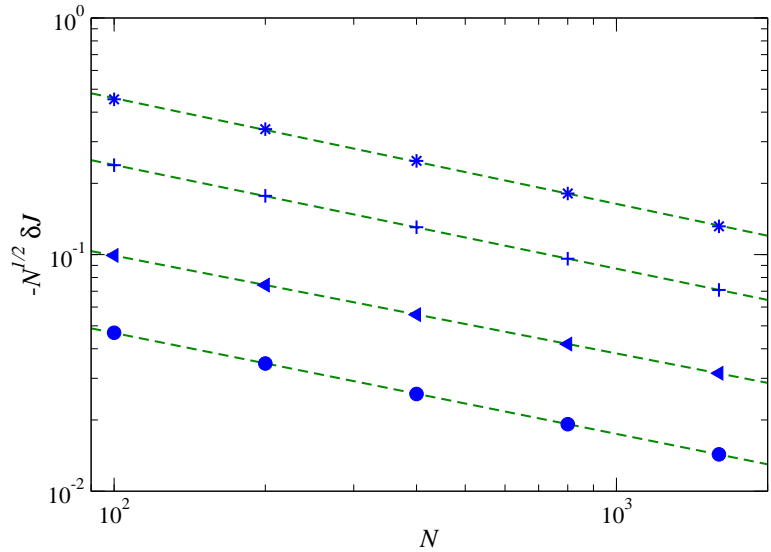}
\caption{Finite size deviation of the heat flux $\delta J \equiv J - J_{num}$,
  rescaled  by $N^{1/2}$, as  a function  of the  size of  the chain  $N$, for
  $\gamma=$  $1$ (circles), $2$  (triangles), $5$  (pluses) and  $10$ (stars),
  other  parameters  as in  the  previous  figure.   The lines  correspond  to
  power-law fits,  from which  we extract that  the corrective terms  scale as
  $-\delta J \sim N^{-\alpha} $  with $\alpha = {0.927}$, ${0.914}$, ${0.939}$
  and ${0.947}$, respectively.
\label{fig:J}}
\end{center}
\end{figure}

As a matter  of fact, the estimation of the  higher-order terms, starting from
the leading corrections  is a highly nontrivial problem,  since such terms are
likely to  be non-analytic in the  smallness parameter $\e$.  This  is seen by
comparing  the  analytical results  and  the  numerical  solutions for  finite
chains. The first  evidence is presented in figure  \ref{fig:T}, where we have
plotted  the  analytical  profile  $T(y)$  and the  numerical  ones  $T_{num}$
computed for chains  of different lengths $N$. From the data  in the inset, we
deduce that $T - T_{num}$  is approximately proportional to $N^{-1/3}$.  While
this confirms  the correctness of  expression \eref{eq:T2}, it  also indicates
that the  leading correction  is of $\Or(\e^{2/3})$.   Nonanalytic corrections
affect also the heat current. This is illustrated in figure \ref{fig:J}, where
we  plot  the  difference  between  the numerical  values  $J_{num}$  and  the
leading-order term $J$, formula \eref{J:8}, for different system sizes $N$ and
for  various  $\gamma$  values.  In  all  cases,  we  see a  clean  power--law
convergence to zero but the value  of the power is systematically smaller than
1, meaning once  again that non-analytic higher-order terms  in $\e$ exist. An
appropriate scheme  for the  treatment of higher-order  terms remains  an open
question.

\ack

We acknowledge useful discussions with L. Delfini and R. Livi.

\section*{References}
\bibliographystyle{unsrt}
\bibliography{dupal}

\end{document}